# Market Equilibrium for Bundle Auctions and the Matching Core of Nonnegative TU Games

By
Somdeb Lahiri.
*Institute for Financial Management and Research*
*24, Kothari Road, Nungambakkam,*
*Chennai- 600 034,*
*Tamil Nadu, India.*
Email:lahiri@ifmr.ac.in
Or **lahiri@webmail.co.za**

## *Abstract*

We discuss bundle auctions within the framework of an integer allocation problem. We show that for multi-unit auctions, of which bundle auctions are a special case, market equilibrium and constrained market equilibrium are equivalent concepts. This equivalence, allows us to obtain a computable necessary and sufficient condition for the existence of constrained market equilibrium for bundle auctions. We use this result to obtain a necessary and sufficient condition for the existence of market equilibrium for multi-unit auctions. After obtaining the induced bundle auction of a nonnegative TU game, we show that the existence of market equilibrium implies the existence of a possibly different market equilibrium as well, which corresponds very naturally to an outcome in the matching core of the TU game. Consequently we show that the matching core of the nonnegative TU game is non-empty if and only if the induced market game has a market equilibrium.

JEL Classifications:

**Introduction**: The formation of productive partnerships and sharing the yield that accrues from it, has been an important concern of both economics and game theory. The theory of matching which originated in the seminal paper of Gale and Shapley (1962), is largely concerned with the formation of stable two agent partnerships. The theory of two-sided matching as discussed in Roth and Sotomayor (1990) that grew out of the work of Gale and Shapley, emphasizes the formation of partnerships between pairs of agents, where each pair consists of agents on two distinct sides of a market. Similarly, the three-sided matching model of Alkan (1988)is concerned with the formation of possible triplets, each triplet comprising agents from three different sets. A related model due to Shapley and Scarf (1974) called the housing market, considers a private ownership economy, where each individual owns exactly one object and what is sought is the existence of an allocation in the core of the economy.

The model discussed here is based on the one proposed by Kaneko and Wooders (1982). They considered a feasible set of coalitions on which a worth function was defined and investigated the problem of there being a non-empty core for such problems. Kaneko and Wooders (1982), called their model a TU partitioning game. Since all coalitions are feasible in our model, we call ours a nonnegative TU game. In a final section of our paper, we discuss how our model can be used to deal with situations represented by TU partitioning games.

Motivated by the work of Eriksson and Karlander (2001) and the literature on coalition formation, we investigate conditions under which there exists an outcome for non-negative TU games satisfying the following conditions:

(a) the realized coalitions form a partition of the set of agents;
(b) the worth of each coalition in the partition is distributed exactly among the agents in the coalition;
(c) no feasible coalition is worth more than the sum of what its members receive.

We call the set of such outcomes, the matching core of the nonnegative TU game. The partition component of an outcome in the matching core is called a matching.

The matching core of a nonnegative TU game is different from the core of a TU game. Necessary and sufficient conditions for the non-emptiness of the core of a TU game, were first obtained by Bondareva (1963) and Scarf (1967). Our analysis and results reported here, are of a very different nature altogether.

The result we seek is a pursuit similar to that of Proposition 2.4 of Eriksson and Karlander (2001). However, our result is different. In effect, we provide a necessary and sufficient condition for the existence of a non-empty matching core, for such problems, when pay-offs are transferable among the agents. In our definition of the matching core, we do not require that the sum of pay-offs to all players be equal to the worth of the grand coalition, even if the latter was a feasible coalition. We require instead, that the sum of pay-offs to players in each coalition that is realized, is equal to the worth of that coalition. The possibility of the grand coalition not being realized is a distinct possibility, and hence "budget balancedness" being superimposed on our solution concept, appears irrelevant. Thus, our matching core is weaker than the core in Eriksson and Karlander (2001).

Moulin (1995), contains a discussion of restrictions in the pattern of coalition formation for games where utilities are transferable. Games of this sort that in addition admit non-empty cores, are called universally stable.

Our approach in this paper is based on results that may be derived for bundle auctions. A bundle auction is one where a non-empty finite set of items is distributed among a non-empty finite set of buyers.

Suppose each agent is a seller of an indivisible item that it owns and each possible coalition is represented by a buyer of such items. A coalition realizes it worth as pay-off if and only if it is able to secure the items that initially belong to its members. Otherwise, the coalition gets zero pay-off. Such a bundle auction is said to be induced by the nonnegative TU game. We show that the matching core of the nonnegative TU game is non-empty if and only if the induced market game has a market equilibrium.

In the next section of this paper we discuss multi-unit auctions and bundle auctions within the framework of an integer allocation problem discussed in Lahiri (2006). We show that for multi-unit auctions of which bundle auctions are a special type, market equilibrium and constrained market equilibrium are equivalent concepts. A constrained market equilibrium, is a market clearing profit maximizing price allocation pair, which arises when buyers are prevented to demand more than the initial endowment of any commodity at any price. Constrained market equilibrium is easier to conceive in the context of a bundle auction. This equivalence, allow us to obtain a computable necessary and sufficient condition for the existence of constrained market equilibrium for bundle auctions. A proof of this result is provided in an appendix to this paper. A significant earlier contribution to this line of research is due to Bikhchandani and Mamer (1997), who showed that market equilibrium for bundle auctions exist if and only if the social value of the optimal allocation is equal to the value of the corresponding relaxed linear program. Other notable existence results of market equilibrium for bundle auctions are due to Keslo and Crawford (1982) and Gul and Stacchetti (1997). Our existence result is different from the ones mentioned above.

For multi-unit auctions, a market equilibrium is no different from a market equilibrium of a bundle auction, where each unit of each good is treated as a different item. Hence, for multi-unit auctions, a necessary and sufficient condition for the existence of market equilibrium follows immediately from the corresponding existence theorem for bundle auctions. Bikhchandani and Ostroy (1998), obtain existence results for what they call the package assignment model that are analogous to the existence result obtained by Bikhchandani and Mamer (1997) for bundle auctions. The package assignment model is precisely what we refer to here as a multi-unit auction.

After obtaining the induced bundle auction of the TU game, we show that the existence of a market equilibrium implies the existence of a possibly different market equilibrium as well, where each coalition either consumes the items initially owned by its members or nothing at all, and the price vector is such that the profit of each coalition is zero. Such a market equilibrium is then shown to correspond very naturally to an outcome in the matching core of the TU game, from which our main result follows.

The interesting thing to note in this context, is that a nonnegative TU game may have an empty matching core, as Example 1 aptly illustrates. In view of the results we derive in this paper, the corresponding induced bundle auction has neither a constrained market equilibrium, nor a market equilibrium.

The results obtained here for nonnegative TU games can be easily applied to situations where certain coalitions are prohibited from being realized, as with TU partitioning games. This can be achieved by setting the worth of a prohibited coalition to be zero. Thus, the implications of our analysis for partitioning games are as valid as they are for nonnegative TU games.

**Multi-unit Auctions and Bundle Auctions**: The model in this section is a variation of the one in Lahiri (2006).

Let $Z = \aleph \cup \{0\}$, where $\aleph$ denotes the set of natural numbers. Let there be $H > 0$ agents and $L+1 > 1$ commodities. The first $L$ commodities are used as inputs to produce the $L+1^{th}$ commodity, which is a numeraire consumption good. Let $w \in Z^L$ denote the aggregate initial endowment of the inputs which is available for distribution among the agents. For $j = 1,\ldots,L$, let $w_j$ denote the aggregate amount of commodity $j$ that is initially available in the economy.

A function f: $Z^L \to \Re_+$ (: the set of non-negative real numbers) is said to be a discrete function.

Each agent i has preferences defined over $Z^L$ which is represented by a discrete production function $f^i$, such that or all i = 1,...,H, $f^i$ is non-decreasing (i.e. for all x,y∈$Z^L$:[x ≥ y] implies [$f^i(x) \geq f^i(y)$]).

The pair <{$f^i$/i= 1,...,H}, w> is called an integer allocation problem.

Our definition of an integer allocation problem here is more general than the one available in Lahiri (2006), since we neither invoke discrete concavity, nor do we invoke any well-behavedness assumption for the individual production functions.

Let e denote the vector in $\Re^L$ all whose coordinates are equal to one and for j = 1,...,L, let $e^j$ denote the vector in $\Re^L$ whose $j^{th}$ coordinate is equal to one and all other coordinates are equal to zero.

For x∈$Z^L$, let C(x) = {y∈$Z^L$/ y ≤ x} and let C*(x) = Convex hull of C(x).

Note that if for a positive integer K, $x^k \in C(e)$ and $t^k \in (0,1)$ for k = 1,...,K, then $\sum_{k=1}^{K} t^k = 1$ and $\sum_{k=1}^{K} t^k x^k = x \in C(e)$ implies $x^k = x$ for k = 1,...,K. If for j∈{1,...,L}, $x^k(j)$ denotes the $j^{th}$ coordinate of $x^k$ and x(j) denotes the $j^{th}$ coordinate of x, then: (a) $0 = \sum_{k=1}^{K} t^k x^k(j)$ if and only if $x^k(j) = 0$ for k = 1,...,K; (b) $1 = \sum_{k=1}^{K} t^k x^k(j)$ if and only if $x^k(j) = 1$ for k = 1,...,K.

Thus, $x^k(j) = x(j)$ for k = 1,...,K.

Thus if f:C(e)→$\Re$, then the function f*:C*(e)→$\Re$ where for y∈C*(e): f*(y) = Max{$\sum_{x \in C(e)} \alpha(x) f(x)$ / $\alpha(x) \geq 0$ for all x∈C(e), $\sum_{x \in C(e)} \alpha(x) = 1$, $\sum_{x \in C(e)} \alpha(x) x = y$} is always concave on C*(e).

An input consumption vector of agent i is denoted by a vector $X^i \in Z^L$.

A price vector p is an element of $\Re_+^L \setminus \{0\}$, where for j = 1,...,L, $p_j$ denotes the price of input j.

At a price vector p, the objective of agent i is to maximize profits:
Maximize [$f^i(X^i) - p^T X^i$]

An allocation is an array X = <$X^i$/ i = 1,...,H> such that $X^i \in Z^L$ for all i = 1,...,H.

Given x∈$Z^L$, let F(x) = { X = <$X^i$/ i = 1,...,H>/ X is an allocation satisfying $\sum_{i=1}^{H} X^i = x$}.

An allocation X is said to be feasible if X∈F(w).

A market equilibrium is a pair <p*,X*> where p* is a price vector, X* is a feasible allocation and for all i = 1,...,H, $X^{*i}$ maximizes profits for agent i.

The function V:$Z^L \to \Re_+$ such that for all x∈$Z^L$: V(x) = Max{$\sum_{i=1}^{H} f^i(X^i)$/ X = <$X^i$/ i = 1,...,H>∈F(x)}, is called the maximum value function.

Since we have assumed that for all i = 1,...,H, $f^i$ is non-decreasing, it must be the case that V is non-decreasing as well (i.e. for all x,y∈$Z^L$:[x ≥ y] implies [V(x) ≥ V(y)]).

A feasible allocation $X^* = \langle X^{*i}/i = 1,\ldots,H\rangle$ is said to be efficient if $\sum_{i=1}^{H} f^i(X^{*i}) = V(w)$.

A constrained market equilibrium is a pair $\langle p^*, X^*\rangle$ where $p^*$ is a price vector, $X^*$ is a feasible allocation such that for all $i = 1,\ldots,H$, $X^{*i}$ solves:
Maximize $[f^i(X^i) - p^T X^i]$
Subject to $X^i \leq w$.
Clearly a market equilibrium is a constrained market equilibrium.
An integer allocation problem $\langle\{f^i/i= 1,\ldots,H\}, w\rangle$ is said to be a multi-unit auction if for all $i = 1,\ldots,H$ and $x \in Z^L$, $f^i(x) = \text{Max}\{f(z)/\ z \in C(y) \cap C(w)\} = \text{Max}\{f(z)/\ z^j \leq \min\{x^j, w^j\}$ for all $j = 1,\ldots,L\}$.
For $x, y \in Z^L$ let $m(x,y)$ be the L-vector whose $j^{th}$ coordinate is $\min\{x^j, y^j\}$.
Thus, if $\langle\{f^i/i= 1,\ldots,H\}, w\rangle$ is a multi-unit auction then for all $i = 1,\ldots,H$ and $x \in Z^L$, $f^i(x) = f^i(m(x,w))$.

Theorem 1: Let $\langle\{f^i/i= 1,\ldots,H\}, w\rangle$ be a multi-unit auction. Then $\langle p^*, X^*\rangle$ is a market equilibrium if and only if it is a constrained market equilibrium.

Proof: We only need to show that if $\langle p^*, X^*\rangle$ is a constrained market equilibrium for the multi unit auction $\langle\{f^i/i= 1,\ldots,H\}, w\rangle$, then it is also a market equilibrium.
Hence suppose $\langle p^*, X^*\rangle$ is a constrained market equilibrium. Let $x \in Z^L$.
If $x \leq w$, then $x = m(x,w)$ and clearly for all $i = 1,\ldots,H$: $f^i(X^{*i}) - p^{*T}X^{*i} \geq f^i(x) - p^{*T}x$.
Thus suppose, $x^j > \min\{x^j, w^j\}$ for some $j = 1,\ldots,L$. Hence for all $i = 1,\ldots,H$: $f^i(X^{*i}) - p^{*T}X^{*i} \geq f^i(m(x,w)) - p^{*T}m(x,w) \geq f^i(x) - p^{*T}x$. Q.E.D.

An multi-unit auction $\langle\{f^i/i= 1,\ldots,H\}, w\rangle$ is said to be a bundle auction if $w = e$.
For $x \in Z^L$, let $e(x) = \sum_{\{j/x_j>0\}} e^j$, where $\sum_{\{j/x_j>0\}} e^j = 0$ if $x = 0$. Thus, $x \geq e(x)$ for all $x \in Z^L$.
Further, $e(x) = x$ if and only $e \geq x$.

Thus, if $\langle\{f^i/i= 1,\ldots,H\}, e\rangle$ is a bundle auction then for all $x \in Z^L$: $m(x,e) = e(x)$.

If $\langle\{f^i/i= 1,\ldots,H\}, e\rangle$ is a bundle auction we represent it simply as $[f^i/i= 1,\ldots,H]$.

It follows from Theorem 1, that for multi-unit auctions in general and bundle auctions in particular, constrained market equilibrium and market equilibrium are identical concepts. Hence the extension of the production functions $f^i$, for $i = 1,\ldots,H$, outside the unit cube is irrelevant for equilibrium analysis in the context of bundle auctions.

A function $f: Z^L \to \Re$ is said to be Weakly Monotonic at $w$ if:
(1) For all $j = 1,\ldots,L$: $f(w + e^j) \geq f(w)$;
(2) $f(w + e) > f(w)$.

An immediate consequence of Theorem 1 above and Theorem 4 of Lahiri (2006) is the following result, which is of considerable interest by itself.

Theorem 2: A necessary and sufficient condition for a bundle auction $[f^i/i= 1,\ldots,H]$ with $V:Z^L \to \Re_+$ weakly monotonic at e, to admit a (constrained) market equilibrium is that: $V(e)$ is the optimal value of the linear programming maximization problem:

Maximize $\sum_{x \in C(2e)} \alpha(x)V(x)$

Subject to $\sum_{x \in C(2e)} \alpha(x)x = e$,

$\sum_{x \in C(2e)} \alpha(x) = 1$,

$\alpha(x) \geq 0$ for all $x \in C(2e)$,

where $C(2e) = \{x \in Z^L / x \leq 2e\}$.

For the sake of completeness an independent proof of Theorem 2, which follows closely the one available in Lahiri (2006) for integer allocation problems (under the additional discrete concavity and well-behavedness assumptions for individual production functions), is provided in an appendix to this paper.

Without the Weak Monotonicity of V at e, we are not able to ensure that the vector of prices is non-zero. We can only guarantee that the vector is non-negative, which is not enough for it to be a price vector.

Given $y \in Z^M$, let $y^- \in Z^M$ be defined thus: $y^-(k) = \min \{y(k), 1\}$ for all $k =1,\ldots,M$. For $i \in \{1,\ldots,H\}$, define $h^i: Z^M \to \Re_+$ as follows: for $y \in Z^M$: $h^i(y) = f^i(x(y^-))$.

Equivalence Theorem For Multi-unit Auctions:
(a) $[h^i/ i = 1,\ldots,H]$ is a bundle auction on M commodities;
(b) $<p,X>$ is a market equilibrium for $<\{f^i/i= 1,\ldots,H\}, w>$ if and only if $<q,Y>$ is a market equilibrium for $[h^i/ i = 1,\ldots,H]$, where for all $j \in \{1,\ldots,L\}$ and $k \in I_j$: $q_k = p_j$ and Y is a feasible allocation for $[h^i/ i = 1,\ldots,H]$ satisfying $x(Y^i) = X^i$ for all $i = 1,\ldots,H$.

Proof: That (a) is true is easily observed. Hence let us prove (b).
Let $<p,X>$ be a market equilibrium for $<\{f^i/i= 1,\ldots,H\}, w>$. Let $q \in \Re_+^M$ and Y be defined as follows: (a) for all $j \in \{1,\ldots,L\}$: $q_k = p_j$ for all $k \in I_j$; (b) For all $j \in \{1,\ldots,L\}$, agent 1 gets the first $X^1$ of the $w^j$ items in $I_j$ at Y, agent 2 gets the next $X^2$ of the remaining items in $I_j$ at Y,… and so on, till agent H gets the last $X^H$ items in $I_j$ at Y.
Clearly $<p,Y>$ is a market equilibrium for $[h^i/ i = 1,\ldots,H]$.
Now, let $<q,Y>$ be a market equilibrium for $[h^i/ i = 1,\ldots,H]$.
If for some $j \in \{1,\ldots,L\}$ with $w^j > 1$, it is the case that two different agents consume two distinct units of j at different prices, then the one who bought it at the higher price, could increase its profit by purchasing the less priced unit instead of the unit it has been allocated. Thus, $<q,Y>$ could not be a market equilibrium.
If for some $j \in \{1,\ldots,L\}$ with $w^j > 1$, it is the case that some agent i consumes all $w^j$ units but at different prices, then on transferring one unit of consumption of the $j^{th}$ good from unit k* to unit k', where k*,k'$\in I_j$ and k* $\neq$ k' in the bundle auction, the decrease in i's profits should be $f^i(x(Y^i)) - f^i(x(Y^i)-e^j) - q^T Y^i + q^T Y^i - q_{k*} + q_{k'} = f^i(x(Y^i)) - f^i(x(Y^i)-e^j) - q_{k*} + q_{k'} \geq 0$. Thus, $q_{k'} - q_{k*} \geq f^i(x(Y^i)) - f^i(x(Y^i)-e^j)$ for all k', k*$\in I_j$ with k' $\neq$ k*. This once again implies that $q_{k'} = q_{k*}$ for all k',k*$\in I_j$. Thus, for all j$\in\{1,\ldots,L\}$: $q_k$ = Min$\{q_{k'}/$ k'$\in I_j\}$ for all k$\in I_j$.

Let p$\in \Re_+^L$, be such that $p_j$ = Min$\{q_k/$ k$\in I_j\}$ for all j = 1,...,L. Let X be the allocation such that for all i = 1,...,H, $X^i = \sum_{k \in I_j} Y^i(k)$.

Thus, <p,X> is a market equilibrium for <$\{f^i/$i= 1,...,H$\}$, w>. Q.E.D.

The maximum value function $V^\#:Z^M \to \Re_+$ for [$h^i$/ i = 1,...,H], is given by $V^\#(y) = V(x(y))$ for all y$\in Z^M$, where V is the maximum value function for the multi-unit auction <$\{f^i/$i= 1,...,H$\}$, w>.

As a consequence of Theorem 2 and the Equivalence Theorem for Multi-unit auctions we get the following result.

Existence of Market Equilibrium Theorem for Multi-unit Auctions:
Let <$\{f^i/$i= 1,...,H$\}$, w> be a multi-unit auction and suppose the maximum value V satisfies Weak Monotonicity at w. Then, <$\{f^i/$i= 1,...,H$\}$, w> has a (constrained) market equilibrium if and only if V(w) is the optimum value of the following linear programming problem:

Maximize $\sum_{x \in C(2w)} \alpha(x)V(x)$

Subject to $\sum_{x \in C(2w)} \alpha(x)x = w$,

$\sum_{x \in C(2w)} \alpha(x) = 1$,

$\alpha(x) \geq 0$ for all x$\in C(2w)$,

where $C(2w) = \{x \in Z^L/ x \leq 2w\}$.

**The Shapley and Shubik (1972) Assignment Game**: The Shapley and Shubik (1972) assignment game is an example of a bundle auction which satisfies the conditions of Theorem 2 and thus admits a market equilibrium.

A bundle auction [$f^i$/ i = 1,...,H] is said to be Shapley and Shubik Assignment Game, if L = H and for all i = 1,...,H: (i) there exists j$\in\{1,\ldots,L\}$ such that $f^i(j) > 0$; (ii) for all x$\in C(e)$, $f^i(x)$ = Max$\{f^i(e^j)/$ $x^j = 1\}$, if x > 0.

Suppose that for some efficient allocation $X^\#$ it is the case that for all i = 1,...,H: $f^i(X^\#) \geq f^i(e^j)$ for all j = 1,...,L. Then clearly $X^{\#i} \in \{e^j/$ j = 1,...,L$\}$. Let p* = (Min $\{f^i(X^{\#i})/$ i = 1,...,H$\})$e. Thus, <p*, $X^\#$> is a market equilibrium.

Hence assume that for no efficient allocation $X^\#$ it is the case that for all i = 1,...,H: $f^i(X^\#) \geq f^i(e^j)$ for all j = 1,...,L.

Thus, V(2e) > V(e), i.e. V satisfies Weak Monotonicity at e.

Let $\alpha(x) \geq 0$ for all x$\in C(2e)$ be such that $\sum_{x \in C(2e)} \alpha(x) = 1$ and $\sum_{x \in C(2e)} \alpha(x)x = e$.

For each $x \in C(2e)$, let $X^x = \langle X^{x,i}/ i = 1,\ldots,H \rangle \in F(x)$, such that $\sum_{i=1}^{H} f^i(X^{x,i}) = V(x)$.

Thus, for all $i \in \{1,\ldots,H\}$, there exists $X^{*x,i} \in \{e^j/ j = 1,\ldots,L\} \cup \{0\}$, such that $X^{*x,i} \leq X^{x,i}$ and $f^i(X^{*x,i}) = f^i(X^{x,i})$.

Thus, $\sum_{x \in C(2e)} \alpha(x)(\sum_{i=1}^{H} X^{*x,i}) \leq e$ and $V(x) = \sum_{i=1}^{H} f^i(X^{*x,i})$ for all $x \in C(2e)$.

Thus, $\sum_{x \in C(2e)} \alpha(x)V(x) = \sum_{x \in C(2e)} \alpha(x)(\sum_{i=1}^{H} f^i(X^{*x,i}))$.

It is well known that $V(e)$ is the optimum value of the following linear programming maximization problem:

Maximize $\sum_{i=1}^{H} \sum_{j=1}^{L} x^i(j) f^i(e^j)$

Subject to $\sum_{j=1}^{L} x^i(j) \leq 1$ for all $i = 1,\ldots,H$,

$\sum_{i=1}^{H} x^i(j) \leq 1$ for all $j = 1,\ldots,L$

$x^i(j) \geq 0$ for all $i = 1,\ldots,H$ and $j = 1,\ldots,L$.

Since the array $\langle \sum_{x \in C(2e)} \alpha(x) X^{*i}(j) / i = 1,\ldots,H$ and $j = 1,\ldots,L \rangle$ is feasible for the above problem, $\sum_{x \in C(2e)} \alpha(x)V(x) = \sum_{x \in C(2e)} \alpha(x)(\sum_{i=1}^{H} f^i(X^{*x,i})) \leq V(e)$.

By Theorem 2, there exists a market equilibrium for the Shapley and Shubik Assignment Game.

**Games with Transferable Utilities**: Given a positive integer $n \geq 3$, and a set of agents $N = \{1,\ldots,n\}$, let $\Pi$ be the set of all non-empty subsets of $N$.

A non-negative TU game (or game with transferable utilities) is a function $v: \Pi \to \Re_+$, such that: (i) For all $i \in N$: $v(\{i\}) = 0$; (ii) $v(S) > 0$ for at least one $S \in \Pi$.

$S \in \Pi$ is said to be a coalition, and $v(S)$ is said to be the "worth" of the coalition $S$.

The reason why we use the prefix "nonnegative" in the above definition is because in general a TU game need not be non-negative. Further, a TU game normally specifies the worth of the empty set to be zero. We define our game on non-empty sets. By itself, requiring the worth of a singleton to be equal to zero, is a harmless normalization. Requiring the worth of at least one coalition to be positive makes the game non-trivial, as will be observed shortly.

A matching is a partition $A$ of $N$, i.e. $A$ is a non-empty collection of mutually disjoint sets in $\Pi$ whose union is $N$.

A pay-off vector is an element of $\Re_+^N$.

An outcome of is a pair $(A,x)$, where $x$ is a pay-off vector and $A$ is a matching.

An outcome $(A,x)$ is said to belong to the matching core of the nonnegative TU game $v$ if:

(1) for all $S \in A$: $\sum_{i \in S} x(i) = v(S)$;

(3) for all $S \in \Pi$: $\sum_{i \in S} x(i) \geq v(S)$.

Let $C(v)$ denote the set of outcomes that belong to the matching core of v.
Since $v(S) > 0$ for at least one coalition S, $(A,x) \in C(v)$ implies $x \neq 0$.
If we had allowed the worth of every coalition to be zero, then for such a game v, $C(v) = \{(A,0)/ A$ is a partition of $N\}$. Assuming that the worth of at least coalition is positive, rules out such trivial possibilities.
The following example due to Ahmet Alkan shows that the matching core of a nonnegative TU game v may be empty.
Example 1 (due to Ahmet Alkan): Let $N = \{1,2,3,4,5\}$. Let $v(S) = 30$ if S has exactly three agents and zero otherwise. Towards a contradiction suppose $(A,x)$ belongs to $C(v)$. If S is a three agent set belonging to A, then at x, at least one member of S, say j, gets at most 10. Since the agents in $N \backslash S$ get zero, the total amount obtained by j and agents in $N \backslash S$ is less than 30, although they form a three-member set.
Thus, every agent in N gets zero at x. Clearly, $(A,x)$ does not belong to $C(v)$.

**Games with Transferable Utilities as Bundle Auctions**: Let $L = n$ and $H = 2^n - 1$. Let $\Pi = \{S_1,\ldots,S_H\}$. For $i \in \{1,\ldots<H\}$, let $v^i: Z^L \to \Re_+$ be defined as follows:
$v^i(x) = v^i(e(x)) = v(S_i)$ if $S_i \subset \{j/ x_j > 0\}$,
          $= 0$, otherwise.
$[v^i/ i = 1,\ldots,H]$ is said to be the bundle auction induced by v.

If $S_i$ is a singleton, then $v^i(x) = 0$ for all $x \in Z^L$.
For $S \in \Pi$, let $e^S$ be defined to be equal to $\sum_{j \in S} e^j$.

Proposition 1: Let $<p^*,X^*>$ be a market equilibrium for $[v^i/ i = 1,\ldots,H]$. Then, there exists a market equilibrium $<p^*,X^\&>$ such that for all $k = 1,\ldots,H$: $X^{\&k} \in \{0, e^{S_k}\}$.

Proof: Let $S = S_i \in \Pi$ and suppose $X^{*i} = e^Q \notin \{0, e^S\}$.
Case 1: S is a proper subset of Q.
Since $v^i(e^Q) - p^{*T}e^Q \geq v^i(e^S) - p^{*T}e^S = v^i(e^Q) - p^{*T}e^S$.
Thus, $0 \geq p^{*T}(e^Q - e^S) = p^{*T}e^{Q\backslash S} = \sum_{j \in Q\backslash S} p^*_j$.

Since $p^* \geq 0$, $p^*_j = 0$ for all $j \in Q\backslash S$.
Let $X^{\#i} = e^S$, $X^{\#k} = X^{*k} + e^j$ if $S_k = \{j\}$ and $j \in Q\backslash S$, $X^{\#k} = X^{*k}$ otherwise.
It is easy to verify that $<p^*,X^\#>$ is a market equilibrium:
$v^i(X^{\#i}) - p^{*T}X^{\#i} = v(S) - \sum_{j \in S} p^*_j = v(S) - \sum_{j \in Q} p^*_j = v^i(X^{*i}) - p^{*T}X^{*i}$;

$v^k(X^{\#k}) - p^{*T}X^{\#k} = v^k(X^{*k} + e^j) - p^{*T}X^{*k} - p^*_j = v^k(X^{*k} + e^j) - p^{*T}X^{*k}$ if $S_k = \{j\}$ and $j \in Q\backslash S$.
The feasibility of $X^\#$ follows from the feasibility of $X^*$ and the fact that $X^\#$ is obtained from $X^*$ by transferring items from S to one or more coalitions outside S.

For j∈Q\S and $S_k$ = {j}, $X^{*k} + e^j$ does not belong to {0, $e^j$} if and only if $X^{*k}$ is not equal to zero. Since $X^{*k}$ cannot be equal to $e^j$, it follows that $|\{i/X^{*k} \notin \{0, e^{S_k}\}\}| > |\{i/X^{\#k} \notin \{0, e^{S_k}\}\}|$.

Case 2: S\Q is non-empty and Q\S is non-empty.
Thus, $v^i(e^Q) = 0$.
Since $-p^{*T}e^Q = v^i(e^Q) - p^{*T}e^Q \geq v^i(0) - p^{*T}0 = 0$ and since $p^* \geq 0$, we get $p_j^* = 0$ for all j∈Q.
Let $X^{\#i} = 0$, $X^{\#k} = X^{*k} + e^j$ if $S_k$ = {j} and j∈Q, $X^{\#k} = X^{*k}$ otherwise.
It is easy to verify that <$p^*, X^\#$> is a market equilibrium:
$v^i(X^{\#i}) - p^{*T}X^{\#i} = 0 = v^i(X^{*i}) - p^{*T}X^{*i}$;
$v^k(X^{\#k}) - p^{*T}X^{\#k} = v^k(X^{*k} + e^j) - p^{*T}X^{*k} - p_j^* = v^k(X^{*k} + e^j) - p^{*T}X^{*k}$ if $S_k$ = {j} and j∈Q.
The feasibility of $X^\#$ follows from the feasibility of $X^*$ and the fact that $X^\#$ is obtained from $X^*$ by transferring items belonging to coalition S to the members of Q.
For j∈Q and $S_k$ = {j}, $X^{*k} + e^j$ does not belong to {0, $e^j$} if and only if $X^{*k}$ is not equal to zero. Since $X^{*k}$ cannot be equal to $e^j$, it follows that $|\{i/X^{*k} \notin \{0, e^{S_k}\}\}| > |\{i/X^{\#k} \notin \{0, e^{S_k}\}\}|$.

Case 3: Q is a non-empty proper subset of S.
Thus, $v^i(e^Q) = 0$.
Since $-p^{*T}e^Q = v^i(e^Q) - p^{*T}e^Q \geq v^i(0) - p^{*T}0 = 0$ and since $p^* \geq 0$, we get $p_j^* = 0$ for all j∈Q.
Let $X^{\#i} = 0$, $X^{\#k} = X^{*k} + e^j$ if $S_k$ = {j} and j∈Q, $X^{\#k} = X^{*k}$ otherwise.
It is easy to verify that <$p^*, X^\#$> is a market equilibrium:
$v^i(X^{\#i}) - p^{*T}X^{\#i} = 0 = v^i(X^{*i}) - p^{*T}X^{*i}$;
$v^k(X^{\#k}) - p^{*T}X^{\#k} = v^k(X^{*k} + e^j) - p^{*T}X^{*k} - p_j^* = v^k(X^{*k} + e^j) - p^{*T}X^{*k}$ if $S_k$ = {j} and j∈Q.
The feasibility of $X^\#$ follows from the feasibility of $X^*$ and the fact that $X^\#$ is obtained from $X^*$ by transferring items belonging to S to the members of Q.
For j∈Q and $S_k$ = {j}, $X^{*k} + e^j$ does not belong to {0, $e^j$} if and only if $X^{*k}$ is not equal to zero. Since $X^{*k}$ cannot be equal to $e^j$, it follows that $|\{i/X^{*k} \notin \{0, e^{S_k}\}\}| > |\{i/X^{\#k} \notin \{0, e^{S_k}\}\}|$.

Thus, in each case we obtain a market equilibrium <$p^*, X^\#$> such that $|\{i/X^{*k} \notin \{0, e^{S_k}\}\}| > |\{i/X^{\#k} \notin \{0, e^{S_k}\}\}|$.
Repeating the process at most finitely many times, we arrive at a market equilibrium <$p^*, X^\&$> such that for all k = 1,…,H: $X^{\&k} \in \{0, e^{S_k}\}$. Q.E.D.

Proposition 2: Let <$p^*, X^*$> be a market equilibrium for [$v^i$/ i = 1,…,H] such that for all i = 1,…,H: $X^{*i} \in \{0, e^{S_i}\}$. Then there exists a market equilbrium <$q^*, X^*$> such that $v^i(X^{*i}) - q^{*T}X^{*i} = 0$ for all i = 1,…,H.

Proof: Since <$p^*, X^*$> is a equilibrium for the induced bundle auction [$v^i$/i = 1,…,H] and $X^{*i} \in \{0, e^{S_i}\}$ for all i = 1,…,H, it must be the case that $\{S_i/ X^{*i} \neq 0\}$ is a partition of N. Further, $v^i(X^{*i}) - p^{*T}X^{*i} \geq 0$ for all i = 1,…,H.

For $i \in \{1,\ldots,H\}$ and $X^{*i} = e^{S_i}$, let $q_j^* = p_j^* + (\frac{v(S_i) - p^{*T} X^{*i}}{|S_i|})$ for all $j \in S_i$.

Thus for $i \in \{1,\ldots,H\}$ and $X^{*i} = e^{S_i}$:
$q^{*T}X^{*i} = q^{*T}e^{S_i} = \sum_{j \in S_i} q_j^* = \sum_{j \in S_i} p_j^* + v(S_i) - p^{*T}X^{*i} = \sum_{j \in S_i} p_j^* + v(S_i) - \sum_{j \in S_i} p_j^* = v(S_i) = v^i(X^{*i})$.

If $X^{*i} = 0$, then $v^i(X^{*i}) - q^{*T}X^{*i} = 0$. Q.E.D.

In view of Proposition 2, we say that a market equilibrium $<p^*,X^*>$ is a **zero-profit market equilibrium** if $v^i(X^{*i}) - q^{*T}X^{*i} = 0$ for all $i = 1,\ldots,H$.

**Market Equilibrium and Matching Cores**: In this section we establish the main consequences of our present investigation.

Theorem 3: $(A,x)$ belongs to $C(v)$ if and only if $<x,X^*>$ is a zero-profit market equilibrium, where for $i = 1,\ldots,H$: $[X^{*i} = e^{S_i}$ if and only if $S_i \in A$; $X^{*i} = 0$ if and only if $S_i \in \Pi \backslash A]$.

Proof: Let $(A,x)$ belong to the matching core of $v$ and $<x,X^*>$ be as defined in the statement of this theorem. Thus, $x \geq 0$ and $x \neq 0$. This implies that $x$ is a price vector. Suppose $S_i \in A$.
Thus, $X^{*i} = e^{S_i}$ and $v^i(X^{*i}) - x^T X^{*i} = v(S_i) - \sum_{j \in S_i} x(j) = 0$. Further, $v^i(e^S) - x^T e^S = v(S_i) - \sum_{j \in S} x(j) \leq v(S_i) - \sum_{j \in S_i} x(j) = 0$ if $S_i \subset S$.

If $S_i \not\subset S$, then $v^i(e^S) - x^T e^S = 0 - \sum_{j \in S} x(j) \leq 0$ and $v^i(0) - x^T 0 = 0$.

Since $\sum_{i=1}^{H} X^{*i} = \sum_{S_i \in A} e^{S_i} = e$, $<x,X^*>$ is a zero profit market equilibrium.

Now suppose $<x,X^*>$ is a zero profit market equilibrium and let $A = \{S_i \in \Pi / X^{*i} = e^{S_i}\}$.
If $v(S_i) - \sum_{j \in S_i} x(j) = v^i(e^{S_i}) - \sum_{j \in S_i} x(j) = 0$ if $S_i \in A$.
If $S_i \notin A$, then $v(S_i) - \sum_{j \in S_i} x(j) = v^i(e^{S_i}) - \sum_{j \in S_i} x(j) \leq v^i(0) - x^T 0 = 0$.

Thus, $(A,x) \in C(v)$. Q.E.D.

The following result follows directly from Propositions 1, 2 and Theorem 3, and is the main consequence of the analysis reported above.

Theorem 4: Let $v$ be a non-negative TU game and $[v^i/ i = 1,\ldots,H]$ be the bundle auction induced by $v$. Then $C(v)$ is non-empty if and only if there exists a market equilibrium for $[v^i/ i = 1,\ldots,H]$.

Let us use Example 1 and Theorem 2 to illustrate Theorem 4. In Example 1, C(v) is empty. Let $[v^i/ i = 1,\ldots,H]$ be the induced bundle auction where $H = 2^5 - 1$ and $L = 5$. In $\Pi$, let $S_1 = \{1,2,3\}$, $S_2 = \{3,4,5\}$, $S_3 = \{1,2,4\}$ and $S_5 = \{5\}$. Let X be the allocation where for i = 1,2,3,4, $X^i = e^{S_i}$; for i > 4, let $X^i = 0$. Clearly $X \in F(2e)$. Thus, $V(2e) \geq \sum_{i=1}^{4} v^i(e^{S_i}) = \sum_{i=1}^{4} v(S_i) = 90$. Further, $V(e) = 30$, and V is Weakly Monotonic at e.

Since $e = \frac{1}{2}2e + \frac{1}{2}0$ and $V(e) = 30 < \frac{1}{2}(90) + \frac{1}{2}(0) \leq \frac{1}{2}V(2e) + \frac{1}{2}V(0)$, by Theorem 2 the induced bundle auction has no market equilibrium, as was to be expected from Theorem 4.

The non-existence of market equilibrium for this induced bundle auction can also be established without using Theorem 2. We then need to use Propositions 1 and 2 instead. Towards a contradiction suppose <p*, X*> is a zero profit market equilibrium such that for all i = 1,...,H: $X^{*i} \in \{0, e^{S_i}\}$. Since a market equilibrium satisfies efficiency, there exists a coalition S containing at least three agents such that S gets $e^S$. Suppose $S = S_i$. Thus, $v^i(e^{S_i}) - \sum_{j \in S_i} p_j^* = 0$. Since, $v^i(e^{S_i}) = 30$, there exists $k \in S_i$ such that $p_k^* > 0$.

If $|S| > 3$, then $v^i(e^{S_i \setminus \{k\}}) - \sum_{j \in S_i \setminus \{k\}} p_j^* = v^i(e^{S_i}) - \sum_{j \in S_i} p_j^* + p_k^* = p_k^* > 0 = v^i(e^{S_i}) - \sum_{j \in S_i} p_j^*$,

contradicting that <p*,X*> is a market equilibrium.

Hence suppose $|S| = 3$. It is easily observed that for $j \in N \setminus S$: $p_j^* = 0$.

Thus, $v^i(e^{(S_i \setminus \{k\}) \cup (N \setminus S)}) - \sum_{j \in (S_i \setminus \{k\}) \cup (N \setminus S)} p_j^* = v^i(e^{S_i}) - \sum_{j \in S_i} p_j^* + p_k^* = = p_k^* > 0 = v^i(e^{S_i}) - \sum_{j \in S_i} p_j^*$,

contradicting that <p*,X*> is a market equilibrium.
The non-existence of a market equilibrium now follows from Propositions 1 and 2.

**Discussion**: It may often be the case that the problem being considered prohibits the formation of certain coalitions. Thus for instance in a two-sided matching model, two or more agents on the same side of the market cannot form a coalition. Our statement of a nonnegative TU game is general enough to accommodate such possibilities.
If S is a coalition which is prohibited then we set its worth v(S) to be equal to zero. Let (A,x) belong to C(v). If S does not belong to A, then there is clearly no problem to be addressed. What if S belongs to A?
If S belongs to A, then x(k) must be equal to zero for all k in S. If instead of (A,x) we considered the outcome (A*,x) where $A^* = (A \setminus \{S\}) \cup \{\{k\}/ k \in S\}$, then it is easily verified that this new outcome belongs to C(v) as well. The difference between A and A* is that all prohibited coalitions in A are replaced by their members. Thus, our model is general enough to cope with the exigencies that arise in matching problems, particularly in the context of markets.

*Acknowledgment*: This problem was indirectly suggested to me by Ahmet Alkan. I would like to thank him for his comments on an earlier version of this paper, and in particular for Example 1.

## *Appendix*

Let $[f^i/\ i = 1,\ldots,H]$ be a bundle auction and let $V:Z^L \to \Re_+$ denote the corresponding maximum value function.

Theorem A.1: Let $X^*$ be a feasible allocation and $p^*$ a price vector. $\langle p^*,X^* \rangle$ is a market equilibrium if and only if the pair $\langle p^*,m^* \rangle$ solves:

Minimize $\sum_{i=1}^{H} m(i) + p^T w$

Subject to $\sum_{i=1}^{H} m(i) + p^T x \geq V(x)$ for all $x \in Z^L$ with $x \leq 2e$, $p \in \Re_+^L$

where $m^* \in \Re^H$ with $m^* = \langle m^*(i)/\ i = 1,..H \rangle$ satisfies $m^*(i) = f^i(X^{*i}) - p^{*T}X^{*i}$ for $i = 1,\ldots,H$.

Proof: Suppose $\langle p^*, X^* \rangle$ is a market equilibrium. Let $x \in Z^L$ and $V(x) = \sum_{i=1}^{H} f^i(X^i)$ where $X = \langle X^i / i = 1,\ldots,H \rangle \in F(x)$.

Thus, for all $i = 1,\ldots,H$: $f^i(X^{*i}) - p^{*T}X^{*i} \geq f^i(X^i) - p^{*T}X^i$.

Summing over i we get: $\sum_{i=1}^{H} m^*(i) + p^{*T}x \geq V(x)$ for all $x \in Z^L$.

Thus, $\langle p^*, m^* \rangle$ satisfies the constraints.

Now, let $\langle p, m \rangle$ satisfy the constraints. Thus, $\sum_{i=1}^{H} m(i) + p^T e \geq V(e)$.

However, since $X^*$ is efficient we get $\sum_{i=1}^{H} m^*(i) + p^{*T}e = V(e)$.

Thus, $\sum_{i=1}^{H} m(i) + p^T e \geq \sum_{i=1}^{H} m^*(i) + p^{*T}e$.

Hence, $\langle p^*, m^* \rangle$ solves the minimization problem.

Now, suppose $\langle p^*, m^* \rangle$ solves the given minimization problem. Towards a contradiction suppose $\langle p^*, X^* \rangle$ is not a market equilibrium. By Theorem 1, $\langle p^*, X^* \rangle$ is not a constrained market equilibrium either. Thus, there exists $i \in H$ and $x \in Z^L$, with $x \leq e$, such that $f^i(x) - p^{*T}x > f^i(X^{*i}) - p^{*T}X^{*i}$.

Thus, $x \leq X^{*i} + e$.

Since $X^* \in F(e)$, $w^* = x + \sum_{k \neq i} X^{*k} \leq X^{*i} + e + \sum_{k \neq i} X^{*k} = e + e = 2e$.

Thus, $V(w^*) \geq f^i(x) + \sum_{k \neq i} f^k(X^{*k})$

$> f^i(X^{*i}) - p^{*T}X^{*i} + p^{*T}x + \sum_{k \neq i} f^k(X^{*k}) - p^{*T}\sum_{k \neq i} X^{*k} + p^{*T}\sum_{k \neq i} X^{*k}$

$= f^i(X^{*i}) - p^{*T}X^{*i} + \sum_{k \neq i} f^k(X^{*k}) - p^{*T}\sum_{k \neq i} X^{*k} + p^{*T}w^*$

$= \sum_{i=1}^{H} m^*(i) + p^{*T}w^*$,

leading to a contradiction.
This establishes the theorem. Q.E.D.

Theorem A.2: There exists a market equilibrium if and only if there exists a price vector $p^*$ such that $V(e) - p^{*T}e \geq V(x) - p^{*T}x$ for all $x \in Z^L$ with $x \leq 2e$.

Proof: Let $\langle p^*, X^* \rangle$ be a market equilibrium. Thus $X^*$ must be efficient, and hence $\sum_{i=1}^{H} f^i(X^{*i}) = V(e)$.

Let $x \in Z^L$ with $x \leq 2e$. By Theorem A.1, $V(e) - p^*e \geq V(x) - p^{*T}x$, since $\sum_{i=1}^{H} X^{*i} = e$.

Now suppose there exists a price vector $p^*$ such that $V(e) - p^{*T}e \geq V(x) - p^{*T}x$ for all $x \in Z^L$ with $x \leq 2e$. Let $X^*$ be an efficient allocation. Since the feasible set is finite such an allocation always exists. Thus, $\sum_{i=1}^{H} f^i(X^{*i}) = V(e)$. Let, $m^* \in \Re^H$ with $m^* = <m^*(i)/\ i = 1,..H>$ satisfying $m^*(i) = f^i(X^{*i}) - p^{*T}X^{*i}$ for $i = 1,...,H$. Since, $\sum_{i=1}^{H} m^*(i) + p^{*T}x = V(e) - p^{*T}e + p^{*T}x \geq V(x)$ for all $x \in Z^L$ with $x \leq 2e$, $<p^*, m^*>$ satisfies the constraints of the linear programming problem in Theorem A.1.

Let $m \in \Re^H$ with $m = <m(i)/\ i = 1,..H>$ satisfy $\sum_{i=1}^{H} m(i) + p^T x \geq V(x)$ for all $x \in Z^L$ with $x \leq 2e$.

Thus, $\sum_{i=1}^{H} m(i) + p^T e \geq V(e) = \sum_{i=1}^{H} m^*(i) + p^{*T} e$.

Thus, $<p^*, m^*>$ solves the linear programming problem in Theorem A.1. By Theorem A.1, $<p^*, X^*>$ is a market equilibrium. Q.E.D.

Observe that the cardinality of $C(2e) = 3^L$.

Lemma A.1: $V(e)$ is the optimal value of the linear programming maximization problem:

Maximize $\sum_{x \in C(2e)} \alpha(x)V(x)$

Subject to $\sum_{x \in C(2e)} \alpha(x)x = e$,

$\sum_{x \in C(2e)} \alpha(x) = 1$,

$\alpha(x) \geq 0$ for all $x \in C(2e)$.

if and only if there exists $p \in \Re^L$ such that $V(e) - p^T e \geq V(x) - p^T x$ for all $x \in C(2e)$.

Proof: Let $C(2e) \setminus \{e\}$ be equal to the set $\{x^k/\ k = 1,...,3^L-1\}$.
Suppose $V(e)$ solves the given linear programming problem. Towards a contradiction suppose there does not exist $p \in \Re^L$ such that $V(e) - p^T e \geq V(x) - p^T x$ for all $x \in C(2e)$.
Hence, there does not exist $\alpha, \beta \in \Re_+^L$ and $\gamma \in \Re_+^{3^L-1}$: $\alpha^T(x^k - e) - \beta^T(x^k - e) - \gamma^k = V(x^k) - V(e)$ for all $k = 1,...,3^L-1$.

By Farkas' Theorem there exists $t \in \Re_+^{3^L-1}$ such that $\sum_{k=1}^{3^L-1} t^k(x^k - e) \leq 0$, $-\sum_{k=1}^{3^L-1} t^k(x^k - e) \leq 0$ and $\sum_{k=1}^{3^L-1} t^k [V(x^k) - V(e)] > 0$.

Thus, $\sum_{k=1}^{3^L-1} t^k > 0$.

Dividing the three inequalities above by $\sum_{k=1}^{3^L-1} t^k$, we get there exists $s \in \Re_+^{3^L-1}$ such that

$\sum_{k=1}^{3^L-1} s^k x^k = e$, $\sum_{k=1}^{3^L-1} s^k = 1$ and $\sum_{k=1}^{3^L-1} s^k V(x^k) > V(e)$, contradicting that V(e) solves the given optimization problem.

Hence, there exists $p \in \Re^L$ such that $V(e) - p^T e \geq V(x) - p^T x$ for all $x \in C(2e)$.

Now suppose that there exists $p \in \Re^L$ such that $V(e) - p^T e \geq V(x) - p^T x$ for all $x \in C(2e)$.

Hence there exists $\alpha, \beta \in \Re_+^L$ and $\gamma \in \Re_+^M$: $\alpha, \beta \in \Re_+^L$ and $\gamma \in \Re_+^{3^L-1}$: $\alpha^T(x^k - e) - \beta^T(x^k - e) - \gamma^k = V(x^k) - V(e)$ for all $k = 1, \ldots, 3^L-1$.

By Farkas' Theorem there does not exist $t \in \Re_+^{3^L-1}$ such that $\sum_{k=1}^{3^L-1} t^k(x^k - e) = 0$, and

$\sum_{k=1}^{3^L-1} t^k [V(x^k) - V(e)] > 0$.

Thus, $[t \in \Re_+^{3^L-1}, t^k \geq 0$ for $k = 1, \ldots, 3^L-1$, $e = \sum_{k=1}^{3^L-1} t^k x^k$, $\sum_{k=1}^{3^L-1} t^k = 1]$ implies $[V(e) \geq \sum_{k=1}^{3^L-1} t^k V(x^k)]$.

Thus, V(e) solves the given optimization problem. Q.E.D.

Lemma A.2: Suppose $V: Z^L \to \Re$ is Weakly Monotonic at e. Then, V(e) is the optimal value of the linear programming maximization problem:

Maximize $\sum_{x \in C(2e)} \alpha(x) V(x)$

Subject to $\sum_{x \in C(2e)} \alpha(x) x = e$,

$\sum_{x \in C(2e)} \alpha(x) = 1$,

$\alpha(x) \geq 0$ for all $x \in C(2e)$.

if and only if there exists $p \in \Re_+^L \setminus \{0\}$ such that $V(e) - p^T e \geq V(x) - p^T x$ for all $x \in C(2e)$.

Proof: By Lemma A.1, V(e) is the optimal value of the linear programming maximization problem in the statement of Lemma A.2, if and only if there exists if and only if there exists $p \in \Re^L$ such that $V(e) - p^T e \geq V(x) - p^T x$ for all $x \in C(2e)$.

Suppose towards a contradiction V(e) is the optimal value of the above linear programming maximization problem, but $p_j < 0$, for some j.

Then, $0 \leq V(e + e^j) - V(e)$ by Weak Monotonicity of V at e and $V(e + e^j) - V(e) \leq p^T(e + e^j - e) = p_j < 0$, leads to a contradiction.

Thus $p \in \Re_+^L$.

If $p = 0$, then $0 < V(2e) - V(w)$ by Weak Monotonicity and $V(2e) - V(e) \leq p^T e = 0$, again leads to a contradiction.

Thus, $p \in \Re_+^L \setminus \{0\}$. Q.E.D.

In view of Lemmas A.1 and A.2 and Theorem A.2, we obtain Theorem 2.